\documentclass[sigconf, nonacm, screen]{acmart}
\usepackage{times}  
\usepackage{array, graphics, graphicx, color, xspace, url, multirow, rotating, epsfig, amsmath, subfig}
\usepackage{wrapfig}
\usepackage{tikz}
\usepackage{booktabs}
\usepackage{epstopdf,subfig}
\usepackage{multicol}
\usepackage[format=plain,labelfont=bf,font=small]{caption}
\usepackage{ragged2e,flushend,enumitem,listings}
\usepackage{acronym} 
\usepackage{breakurl}
\usepackage{hyperref, cleveref}
\usepackage{multirow}
\usepackage[frozencache,cachedir=.]{minted}
\usepackage{appendix}
\usepackage[compact,small]{titlesec}
\usepackage[para]{footmisc}
\usepackage{comment}
\usepackage{paralist}
\usepackage{paracol}

\hypersetup{pdfstartview=FitH,pdfpagelayout=SinglePage}

\newcommand\paragraphb[1]{\noindent{\bf{#1}}}
\newcommand\paragraphi[1]{\noindent\emph{#1}}
\newcommand\pb[1]{\paragraphb{#1}}
\renewcommand\pi[1]{\paragraphi{#1}}

\newcommand{\bi}{\begin{itemize}}
\newcommand{\ei}{\end{itemize}}

\newcommand{\eg}{\emph{e.g.,}\xspace}

\newcommand{\eat}[1]{}

\newcommand{\allnotes}[1]{}
\renewcommand{\allnotes}[1]{\textit{#1}} %

\hypersetup{                    %
  colorlinks,
  linkcolor={green!80!black},
  citecolor={red!70!black},
  urlcolor={blue!70!black}
}

\let\svthefootnote\thefootnote
\newcommand\freefootnote[1]{%
  \let\thefootnote\relax%
  \footnotetext{#1}%
  \let\thefootnote\svthefootnote%
}

\graphicspath{{./dia/}}

\newcommand\vldbpagestyle{plain} 

\definecolor{codegray}{rgb}{0.5,0.5,0.5}
\lstdefinestyle{bashStyle}{
    commentstyle=\color{blue},
    numberstyle=\tiny\color{codegray},
    basicstyle=\ttfamily\scriptsize,
    breakatwhitespace=false,         
    breaklines=true,                 
    captionpos=b,                    
    keepspaces=true,                 
    numbers=left,                    
    numbersep=5pt,                  
    showspaces=false,                
    showstringspaces=false,
    showtabs=false,                  
    tabsize=2,
    keywordstyle=\bfseries,
}

\setminted{fontsize=\footnotesize, xleftmargin=0.22in,linenos}

\begin{document}
\title{Compound Schema Registry (Extended Abstract)}

\author{Silvery D. Fu$^{1,2}$, Xuewei Chen$^{1}$}
\affiliation{$^{1}$UC Berkeley, $^{2}$System Design Studio}

\begin{abstract}
\vspace{-0.1in}
Schema evolution is the process of modifying a database system's schema to maintain compatibility with existing data~\cite{kafka-se, databricks-se, hudi-se, vldb12-auto-se}. It allows data producers to update schemas while ensuring they remain compatible with the ones used by downstream consumers. For example, a producer might add a new \texttt{timestamp} field that does not disrupt existing consumers unprepared for this change. 

A schema registry~\cite{kafka-schema-registry,sharma2022schema} is a common approach aiming to address the challenges of schema evolution, especially for real-time data streaming. It serves as a centralized repository to store, manage, validate, and ensure the compatibility of schemas. The registry facilitates communication between producers and consumers through a well-defined data contract encapsulated within a schema. It controls schema evolution through clear and explicit compatibility rules, ensuring that all participants adhere to established standards. The registry optimizes data transmission by using schema IDs instead of full schema definitions. At runtime, the schema registry dynamically resolves these IDs to their corresponding schemas, enabling systems to correctly interpret incoming data streams and integrate schema changes without interruptions.

However, existing schema registries can typically manage only simple modifications to schemas, such as adding or removing fields. More complex \textbf{syntactic} alterations, such as renaming fields, changing data types, or modifying units and scaling, are generally considered breaking changes. These changes can lead to application downtime, requiring a human in the loop to write schema matching and mapping code at the application level to restore compatibility and carefully manage the migration. For instance, in a Kafka ecosystem that includes a data consumer, producer, and schema registry~\cite{kafka-schema-registry}, developers responsible for the consumer application must be notified to update their code before the producer makes any changes to field names or types. Such coordination is crucial to ensure that the consumer continues to receive data correctly. This process can be tedious and often prevents scenarios such as zero-downtime upgrades; it also limits the ability of applications to access real-time data from data sources with previously unknown or changing schemas.

To this end, we propose \textit{generalizing} schema evolution to accommodate a broader range of schema syntax changes. With generalized schema evolution (GSE), as long as the \textbf{semantics} of two fields or schemas remain equivalent or compatible---as determined by the data consumer---data streams will continue uninterrupted when the data producer evolves the schema. We argue that to realize GSE, the schema registry should transform into a compound AI system~\cite{compound-ai-blog}. Our insight is that Large Language Models (LLMs), with their capability to understand data semantics, can significantly improve how schema changes are managed and streamline the \emph{schema mapping} between different schema versions. For example, consider two versions of motion sensor schemas illustrated in Fig.~\ref{code:mapping}. Our approach would enable the automatic mapping of data from version v2 to version v1, allowing data consumers operating under the v1 schema to continue accessing data produced under v2.

We present a design and a prototype for \textit{compound schema registry} to support GSE, which aims to address three key requirements: \textbf{(1) Accurate:} The mappings across schema versions must be precise, ensuring correct generation and application of transformations to fields and values within the schema to the data records. \textbf{(2) Fast and efficient:} Rather than using LLMs to directly translate each data record---a process that can be inefficient and slow due to frequent model calls---we should generate schema mappings and translate them into dataflow operations implemented on the data path (\eg at the data consumer, within the message broker, or integrated into the data pipeline). This approach of \textit{generating off-path code for on-path execution} not only ensures high accuracy but also improves efficiency. \textbf{(3) Transparent:} The mapping process and its outputs should be straightforward and easily verifiable for correctness, avoiding opaque operations (\eg hidden within a single model call). Moreover, we advocate for creating an \textit{intermediate representation} for schema mappings that are independent of specific dataflow languages. This approach simplifies the generation process by avoiding any unnecessary intricacies of individual language syntax, while enabling the reuse of mappings across different data processing engines and platforms. 

\begin{figure}
\centering
\footnotesize
\begin{minipage}{.21\textwidth}
\begin{minted}{yaml}
kind: "Motion sensor"
name: "v1"
description: "Philips Hue"
fields:
  - name: "motion"
    type: "boolean"
    description: >
      True if motion 
      is detected.
    required: true
    
  - name: "enabled"
    type: "boolean"
    description: >
      True when the sensor 
      is activated, false 
      when deactivated.
    required: true

  - name: "sensitivity"
    type: "integer"
    description: >
      Motion sensitivity 
    default: 2
    min: 0
    max: 4
\end{minted}
\end{minipage}
\begin{minipage}{.21\textwidth}
\begin{minted}{yaml}
kind: "Motion sensor"
version: "v2"
description: "Vivint"
fields:
  - name: "triggered"
    type: "boolean"
    description: >
      Indicates whether the 
      sensor has been 
      triggered.
    required: true

  - name: "enabled"
    type: "boolean"
    description: >
      Indicates whether the 
      motion sensor is enabled 
      (True) or bypassed (False).
    required: true

  - name: "battery_percentage"
    type: "integer"
    description: >
      Measures the current battery 
      level of the motion sensor.
    required: true
\end{minted}
\end{minipage}
\caption{Example schemas for motion sensor data (v1 and v2).}
\label{code:motion}
\vspace{-0.25in}
\end{figure}

To meet the above requirements, we propose a \textbf{task-specific language} called Schema Transformation Language (STL), for generating schema mappings as an intermediate representation (IR), instead of directly generating the dataflow operators from the given source and target schemas. The language defines a collection of schema mapping commands, as detailed in Table~\ref{tab:stl}. These include (i) \textit{schema matching} commands for assessing compatibility between schemas, (ii) \textit{field transformation} commands for directly modifying schema fields such as adding, deleting, or renaming them, and (iii) \textit{value transformation} commands for converting field values to comply with new schema specifications. Each command handles a specific sub-task of schema mapping. At runtime, the schema registry uses STL as part of the prompt to invoke an LLM, where each command is defined as a function, \eg using the OpenAI function calling interface in our prototype, along with the two versions of the schemas to be mapped. The LLM then generates schema mappings as STL commands, as depicted in Fig.~\ref{code:mapping}. Subsequently, an \textit{assembler} translates these STL commands into the corresponding dataflow operations using the dataflow language of the target platform, which can then be patched or installed on the data pipeline.

\begin{table*}
\centering
\renewcommand{\arraystretch}{1.2}
\begin{tabular}{ | m{2.7cm} | m{3cm}| m{11cm} | }
    \hline
    \textbf{Command class} & \textbf{Command name} & \textbf{Description} \\
    \hline
    \multirow{1}{*}{Schema matching} & MATCH & Used to determine whether the source and target schemas correspond to the same entity; if they match, the schema mapping will continue; otherwise, it will abort.\\ 
    \hline
    \multirow{7}{*}{Field transformation} & COPY & Directly copies data from the source field to the target field without any transformation.  \\ \cline{2-3}
    & ADD & Inserts a new field into the target schema that does not exist in the source schema. \\ \cline{2-3}
    & CAST & Converts the data type of the source field to match the expected type of the target field. \\ \cline{2-3}
    & DELETE & Removes the field from the source schema when it is not required in the target schema. \\ \cline{2-3}
    & RENAME & Changes the name of the source field to match the name of the target schema. \\ \cline{2-3}
    & DEFAULT & Assigns a predefined default value to a target field when data is unavailable or null. \\ \cline{2-3}
    & MISSING & Used when no appropriate mapping exists to map the source field to a target field, implying a schema mapping failure. \\
    \hline
    \multirow{5}{*}{Value transformation} 
    & SCALE & Adjusts the numerical values in the source field by a specified factor for the target field. \\ \cline{2-3}
    & SHIFT & Modifies the values in the source field by adding or subtracting a constant value. \\ \cline{2-3}
    & LINK & Establishes a correspondence between values in the source field and defined values in the target field, used for fields with enum type. \\ \cline{2-3}
    & GEN & Generates a transformation function that defines how to convert values from the source field to fit the target field's requirements.\\ \cline{2-3}
    & APPLY & Applies a transformation function, either generated or predefined by the developer, to the value of a source field to derive the value of the target field.\\ \cline{2-3}
    \hline
\end{tabular}
\caption{Key commands in Schema Transformation Language (STL) of the compound schema registry.}
\label{tab:stl}
\end{table*}

\begin{figure*}[h]
\vspace{-0.2in}
\centering
\RecustomVerbatimEnvironment{Verbatim}{BVerbatim}{}
\begin{minted}{yaml}
{from: triggered, to: motion, transformation: RENAME triggered TO motion}
{from: battery_percentage, to: None, transformation: DELETE battery_percentage}
{from: None, to: sensitivity, transformation: ADD sensitivity TYPE integer}
{from: sensitivity, to: sensitivity, transformation: DEFAULT sensitivity TO 2}
{from: enabled, to: enabled, transformation: COPY}
\end{minted}
\vspace{-0.1in}
\caption{Generated mappings for motion sensor schema v1 and v2.}
\vspace{-0.05in}
\label{code:mapping}
\end{figure*}

\begin{table}
\centering
\footnotesize
\begin{tabular}{|c|c|cc|cc|cc|}
\hline
\multirow{2}{*}{\textbf{\begin{tabular}[c]{@{}c@{}}Source \\ schema\end{tabular}}} & \multirow{2}{*}{\textbf{\begin{tabular}[c]{@{}c@{}}Target \\ schema\end{tabular}}} & \multicolumn{2}{c|}{\textbf{Precision}} & \multicolumn{2}{c|}{\textbf{Recall}} & \multicolumn{2}{c|}{\textbf{F1}} \\ \cline{3-8} 
 &  & \multicolumn{1}{c|}{\textbf{STL}} & Base & \multicolumn{1}{c|}{\textbf{STL}} & Base & \multicolumn{1}{c|}{\textbf{STL}} & Base \\ \hline
Philips Hue & Vivint & \multicolumn{1}{c|}{\textbf{0.91}} & 0.73 & \multicolumn{1}{c|}{\textbf{0.98}} & 0.83 & \multicolumn{1}{c|}{\textbf{0.94}} & 0.78 \\ \hline
SimpliSafe & Vivint & \multicolumn{1}{c|}{\textbf{1}} & 0.2 & \multicolumn{1}{c|}{\textbf{0.8}} & 0.2 & \multicolumn{1}{c|}{\textbf{0.89}} & 0.2 \\ \hline
SimpliSafe & Philips Hue & \multicolumn{1}{c|}{\textbf{1}} & 0.8 & \multicolumn{1}{c|}{\textbf{0.9}} & 0.67 & \multicolumn{1}{c|}{\textbf{0.95}} & 0.72 \\ \hline
\end{tabular}
\caption{Accuracy of evolving schema with STL and direct model call.}
\label{tab:f1}
\vspace{-0.37in}
\end{table}

Our initial results suggest promising improvements in schema mapping accuracy with STL compared to generating dataflow operators directly using an LLM. For example, when applied to real-world IoT device schemas and schema evolution scenarios, the STL approach can significantly improve the average F1 score—measured based on the precision and recall of generating the correct mappings—from 78\% to 94\% across runs for the example schemas, as shown in Table~\ref{tab:f1}. This is because STL: (1) breaks down the schema mapping task into smaller, specific sub-tasks (\eg field transformations to value transformations for each field), and (2) separates mapping generation from dataflow generation so that each step can be performed more easily. With better per-STL-command prompt engineering, this approach could achieve even higher mapping accuracy. Further, we found that the quality of schema definitions (\eg how concise each field explanation is) plays an important role in mapping accuracy. We assume the schema definitions are given or can be extracted automatically in a separate process~\cite{cidr13-tamer}, which itself can also be performed through a compound AI approach. An interesting question is how we can co-design the schema extraction, mapping, and evolution processes.

We are extending the prototype to handle schema evolution across different target platforms and evaluating it using various datasets~\cite{moller2020evobench,de2017zero}, while comparing it with prior approaches~\cite{vldb12-auto-se,arxiv24-seed,icde23-plm-schema-match}. Our codebase will be made available at \url{https://llmint.org}.

\pb{Design Pattern: Task-Specific Language / IR.} We propose extending the discussed design pattern beyond schema evolution by employing LLMs to generate messages in a task-specific language for broader applications. Within this framework, each command is clearly defined to handle a specific sub-task, with predefined templates for inputs and outputs. Such unambiguous and modular task specification can also make the output verifiable and task execution debuggable. This approach can deliver more \emph{general and reliable} LLM-based automation across various domains, such as workflow automation, data automation, and decision support systems.

\end{abstract}

\maketitle

\pagestyle{\vldbpagestyle}

\vspace{0.08in}
\bibliographystyle{ACM-Reference-Format}
\bibliography{llmint}

\end{document}